\documentclass[reprint]{revtex4-2}
\usepackage{amssymb,amsmath,amsthm}
\usepackage{graphicx}
\usepackage{bm,bbm}

\usepackage[bookmarks=false]{hyperref}
\hypersetup{colorlinks=true,citecolor=blue,linkcolor=red,%
urlcolor=blue,pdfstartview=FitH,bookmarksopen=true}

\usepackage[T1]{fontenc}
\usepackage[osf,sc]{mathpazo}
\usepackage{dsfont}

\newcommand{\ket}[1]{\mbox{$ | #1 \rangle $}}

\newcommand{\tr}{\mathrm{tr}}
\newcommand{\cP}{{\cal P}}

\newcommand{\cN}{\mathcal{N}}

\newcommand{\trans}{^{\mathsf{T}}}

\DeclareMathAlphabet{\vecfont}{OT1}{cmr}{bx}{it}
\DeclareMathAlphabet{\dyadfont}{OT1}{cmss}{bx}{n}
\renewcommand{\vec}[1]{\vecfont{#1}}
\newcommand{\column}[2][c]{{\left(\begin{array}{#1} #2 \end{array}\right)}}

\DeclareMathAlphabet{\vecfont}{OT1}{cmr}{bx}{it}
\DeclareMathAlphabet{\dyadfont}{OT1}{cmss}{bx}{n}
\renewcommand{\vec}[1]{\vecfont{#1}}

\newcommand{\symvec}[1]{\boldsymbol{#1}}

\newtheoremstyle{note}
  {\topsep/2}              	
  {\topsep/2}            	
  {}                        
  {\parindent}             	
  {\itshape}                
  {.---}                    
  {0pt}                     
  {\thmname{#1}\thmnumber{ \itshape#2}\thmnote{ (#3)}} 

\theoremstyle{definition}

\theoremstyle{remark}

\usepackage{algorithm}

\usepackage{mdframed}

\newenvironment{inlinealgorithm}[1]{%
  \refstepcounter{algorithm}
  \par\medskip
  \noindent
  \hrule height 1.2pt
  \vspace{2pt}
  \centerline{\textbf{Algorithm: #1}}
  \vspace{2pt}
  \hrule height 0.6pt
  \vspace{3pt}
  \small
}{%
  \vspace{3pt}
  \hrule height 0.6pt
  \medskip
}
\usepackage{float}

\usepackage{algorithmicx}
\usepackage{algpseudocode}

\begin{document}
\title{Error-mitigated quantum state tomography using neural networks}

\author{Yixuan Hu}
\affiliation{Key Laboratory of Advanced Optoelectronic Quantum Architecture and Measurement (MOE), School of Physics, Beijing Institute of Technology, Beijing 100081, China}

\author{Mengru Ma}
\affiliation{Key Laboratory of Advanced Optoelectronic Quantum Architecture and Measurement (MOE), School of Physics, Beijing Institute of Technology, Beijing 100081, China}

\author{Jiangwei Shang}
\email{jiangwei.shang@bit.edu.cn}
\affiliation{Key Laboratory of Advanced Optoelectronic Quantum Architecture and Measurement (MOE), School of Physics, Beijing Institute of Technology, Beijing 100081, China}

\date{\today}
\begin{abstract}
The reliable characterization of quantum states is a fundamental task in quantum information science.
For this purpose, quantum state tomography provides a standard framework for reconstructing quantum states from measurement data, yet it is often degraded by experimental noise.
Mitigating such noise is therefore essential for the accurate estimation of the states in realistic settings.
In this work, we propose a scalable tomography method based on multilayer perceptron networks that mitigate unknown noise through supervised learning.
This approach is data-driven and thus does not rely on explicit assumptions about the noise model or measurement, making it readily extendable to general quantum systems.
Numerical simulations, ranging from special pure states to random mixed states, demonstrate that the proposed method effectively mitigates noise across a broad range of scenarios, compared with the case without mitigation.
\end{abstract}

\maketitle
%

\section{Introduction}
Characterizing quantum states is a fundamental task in quantum information theory and experiments.
The standard methodology of quantum state tomography (QST) provides a general and powerful technique to characterize unknown quantum states through measurements \cite{wiseman_milburn_2009, Jezek.etal2003}.
While alternative state characterization approaches have also been proposed, such as quantum state verification  \cite{Morimae.etal2017, Pallister.etal2018, Takeuchi.Morimae2018, Yu.etal2019} and direct fidelity estimation \cite{Flammia.Liu2011}.
However, these methods typically rely on specific prior knowledge about the target state or provide only limited information.
In contrast, QST enables full state reconstruction without requiring any prior information through informationally complete measurements on quantum ensembles.
As a result, QST plays a key role in characterizing states, especially those within quantum circuits or quantum computers, and has thus attracted significant research interests \cite{QCQI2010, gisin.etal2007, QSE2004}.
A wide range of tomographic techniques has been developed, such as least squares inversion \cite{Opatrny.eral1997}, Bayesian tomography \cite{Huszar.etal2012, Blume-Kohout_2010}, maximum likelihood estimation \cite{Jezek.etal2003, Rehacek.etal2001}, linear regression estimation \cite{Qi.etal2013, Qi.etal2017}, and convex optimization \cite{Ingrid2022}.

In realistic experimental settings, however, measurement imperfections and environmental noise inevitably contaminate the collected data, leading to biased or inaccurate reconstructions.
Noise mitigation is thus a critical challenge in both quantum information theory and experiment.
One straightforward approach is to explicitly characterize the noise via process tomography \cite{QCQI2010} , but this approach requires additional measurements and assumes noise stability.
Another idea is compressed sensing \cite{Gross.etal2010, Shabani.etal2011, Sherbert.etal2022}, which mitigates noise without extra measurements or explicit noise characterization.
However, compressed sensing relies on a low-rank assumption, making it unsuitable for general cases.
Therefore, limitations like these have motivated growing interests in machine learning techniques, particularly neural networks, which have demonstrated superior performance across a variety of quantum information tasks \cite{lennon.etal2019, Fosel.etal2018, Huang.etal2020, ma2022compression}.

Neural networks, inspired by biological neurons \cite{Rosenblatt1961}, are universal function approximators \cite{Paul1982}.
With the development of multilayer perceptrons (MLPs) and the feedforward-backpropagation algorithm, neurons become neural networks that can approximate general nonlinear mappings \cite{cybenko_approximation_1989}.
However, training deep networks remains difficult in practice, until architectural and algorithmic innovations, such as deep belief networks \cite{Hinton.etal2006-b}, laid the groundwork for modern deep learning \cite{lecun.etal2015}.
Based on this model, specialized architectures further enhance their capabilities, for example, the convolution layers for image sensing \cite{Shelhamer.etal2017}, the diffusion model for image generation \cite{Ho.etal2020}, and attention for natural language processing \cite{vaswani2023attentionneed}.

In the context of QST, neural-network-based approaches have gained increasing attraction in recent years. 
For example, various architectures have been adapted for this task, including the restricted Boltzmann machine \cite{Torlai.etal2018}, conditional generative adversarial networks \cite{Ahmed.etal2021}, and MLPs \cite{Xin.etal2019}.
These models are designed for different tasks and focuses, such as reducing computational complexity \cite{Xu.etal2018, Torlai.etal2018}, improving data efficiency in training \cite{palmieri2023enhancing}, making the neural network more suitable for large systems or reducing the need for measurements \cite{Gaikwad.etal2024, quek.etal2021}.
Importantly, neural networks have also demonstrated a strong potential to mitigate the effects of noise on quantum tomography, including Gaussian and uniform noise models \cite{Tolai.etal2020, ma2021comparative}.
These results suggest that neural networks have the potential to serve as flexible and powerful tools for noise-mitigated quantum state reconstruction.

Our analysis in this work further demonstrates the potential of neural networks to mitigate the effects of unpredictable noise in quantum state tomography.
In contrast to previous work, the noise considered here is not restricted to specific parametric models and may even vary across states.
Using multilayer perceptrons trained via supervised learning, noise effects are learned directly from data and mitigated without explicit noise modeling.
Furthermore, for certain classes of states, the proposed approach can be applied with informationally incomplete measurements, thus reducing the complexity and enabling tomography for large-scale quantum systems.

The outline of this work is as follows.
In Sec.~\ref{Sec:Str}, we introduce the general structure and algorithm of the neural-network-based quantum state tomography.
Section~\ref{Sec:App} presents a detailed analysis of the neural-network approach across various application scenarios.
Finally, we summarize our findings and discuss future perspectives in Sec.~\ref{Sec:Sum}.

\section{Neural-network-based quantum state tomography}\label{Sec:Str}
The neural-network-based quantum state tomography can be summarized briefly as the following.
Within the framework, the input and output of the neural network are specified firstly.
Then, a suitable neural network model is selected that is simple and capable of representing the required mapping.
Notably, no assumptions are imposed on the structure of the noise nor on the detailed form of the measurement model, making it a general method that would fit any experimental requirements.
The neural network does not acquire prior information in advance, but only learns the state and noise from the data it has been given.
The subsequent statements about noise and measurement are provided as an example.

\begin{figure}[t]
    \centering
    \includegraphics[width=0.9\columnwidth]{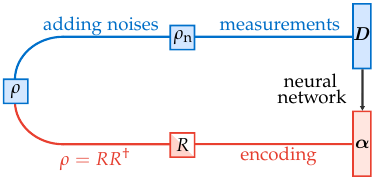}
    \caption{Schematic illustration of the neural-network-based quantum state tomography.
    Starting from a noise-free state $\rho$, two parallel procedures are involved:
   (1) The blue path represents the physical processing, where noisy channels transform $\rho$ into $\rho_n$ followed by measurements yielding the data vector $\vec{D}$;
   (2) The red path denotes the numerical processing, where a Cholesky decomposition produces a lower triangular matrix $R$ which is subsequently reshaped and encoded into the parameter vector $\symvec{\alpha}$.}
    \label{fig:procedure}
\end{figure}
\subsection{General framework}
The general framework is schematically illustrated in Fig.~\ref{fig:procedure}.
The basic procedure is to measure an ideal noise-free state $\rho$ using a measurement set $M$.
All experimental noises in this procedure are assumed to be incorporated into the state, producing a noisy state $\rho_n$, while the measurement process itself is treated as noise-free.
These noises vary randomly and are uncorrelated across different states $\rho$, but remain stable during individual QST processes.
As for the measurement set, it is taken to be informationally complete, and the resulting measurement statistics are collected in the data vector $\vec{D}$.
The reconstruction task is therefore reduced to learning a functional mapping from the measurement outcomes $\vec{D}$ to the underlying quantum state $\rho$.

Neural networks, however, do not inherently guarantee that the output satisfies the physical constraints of the density matrix.
Furthermore, the presence of redundant degrees of freedom, such as the phases, should be eliminated. 
To address these challenges, the lower branch of Fig.~\ref{fig:procedure} is designed as an encoding–decoding process.
Firstly, the physical constraints are enforced by the Cholesky decomposition \cite{ACD1990}, such that \begin{align}
\rho = RR^\dagger\,,
\end{align}
with $R$ being a lower triangular matrix.
Only the independent elements of $R$ are retained and arranged into a real-valued parameter vector $\symvec{\alpha}$, excluding all structurally constrained zero entries.
During decoding, normalization is applied by
${\rho' = RR^\dagger / \tr(RR^\dagger)}$,
thereby ensuring $\rho'$ strictly satisfies the requirements of the density matrix.
This representation not only reduces redundancy, but also improves the computational efficiency and preserves the physical validity.

To further improve the reconstruction fidelity, an additional encoding–decoding scheme is introduced.
This scheme is conceptually inspired by the one-hot representations commonly used in machine learning \cite{HARRIS2016108}.
However, conventional one-hot encoding is unsuitable for continuous variables, here we propose a modified scheme:
\begin{inlinealgorithm}{One-hot inspired encoding}
\begin{algorithmic}
  \Require
 A real-valued parameter vector $\symvec{\alpha}$ of length $k$, the number of sectors $N_{\mathrm{sec}}$, and the empirically determined bounds $\alpha_{\mathrm{max}}$ and $\alpha_{\mathrm{min}}$ for each component $\alpha_i$.
  \Ensure
  An encoded representation $\vec{Y}$ corresponding to $\symvec{\alpha}$.
\Statex
  \Function{One-hot Inspired Encoding}{$\symvec{\alpha}$}
  \State Initialize $\vec{Y}$ as a sparse vector with all entries set to zero.
  \For{$i=0$ to $k-1$} \Comment{index started from $0$}
  \State Compute sector width ${d=({\alpha_{i,\mathrm{max}}-\alpha_{i,\mathrm{min}}})/ N_{\mathrm{sec}}}$ 
  \State Compute normalized position ${y_{\mathrm{int}}=({\alpha_i-\alpha_{i,\mathrm{min}}})/d}$
  \State Determine the sector index $\mathrm{ind}=\lfloor y_{\mathrm{int}}\rfloor$ \Comment{rounding down to get index}
  \State $Y_{i,\mathrm{ind}}$=$(\alpha_i-\alpha_{i,\mathrm{min}}-\mathrm{ind}\cdot d)/d$
  \State $Y_{i,\mathrm{ind}+1}$=$1-Y_{i,\mathrm{ind}}$
  \EndFor
  \EndFunction
\end{algorithmic}
\end{inlinealgorithm}
Each component $\alpha_i$ of the parameter vector $\symvec{\alpha}$ is assumed to lie within a finite interval $[\alpha_{i,\min}, \alpha_{i,\max}]$, which is empirically determined before training.
These intervals are divided into $N_{\mathrm{sec}}$ uniform sectors, thereby discretizing the continuous parameter space.
Rather than assigning $\alpha_i$ to a single discrete value, the encoding represents $\alpha_i$ as a convex combination of two neighboring sector endpoints.
After encoding, the vector $\vec{Y}$ is normalized and interpreted as a probability distribution.
Such characteristics are known to be able to improve the numerical stability and training efficiency in neural networks \cite{huang2020_normalization}.

\subsection{The neural-network architecture}
\begin{figure}[t]
    \centering
    \includegraphics[width=0.95\columnwidth]{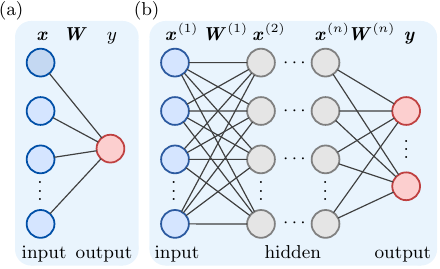}
    \caption{
    Neural network structures: 
    (a) Perceptron model with direct input-output connections, but limited to linear mappings.
    (b) Multilayer perceptron (MLP) with hidden layers, enabling approximation of complex functions.
    The data is feedforward from input to output, while nodes in the middle do not serve as output, turning out to be hidden layers.
    Properly designed MLPs can universally approximate the mappings from noisy measurements to noise-free states.
    }
    \label{fig:nn}
\end{figure}

Within the proposed framework, quantum state tomography is reformulated as a supervised learning problem.
The objective here is to infer the parameter vector $\symvec{\alpha}$ from the measurement data vector $\vec{D}$.
Neural networks serve as universal function approximators \cite{cybenko_approximation_1989}, enabling the representation of highly nonlinear mappings without requiring explicit prior knowledge of their analytical form.
The network architectures used in this work are illustrated in Fig.~\ref{fig:nn}.

The most elementary architecture employed is the perceptron model, shown in Fig.~\ref{fig:nn}(a).
Originally inspired by biological neurons \cite{Rosenblatt1961}, the perceptron serves as a fundamental computational unit in neural networks.
For input ${\vec{x}=\column[cccc]{x_1 &x_2 &\cdots &x_k} \trans}$, and weights ${\vec{W}=\column[cccc]{W_1&W_2&\cdots& W_k}\trans}$, the output is given by 
\begin{align}
    y=f\bigl(\vec{W}\trans \vec{x}\bigr)\,,
\end{align}
where $f$ denotes the activation function.
In this work, the leaky rectified linear unit (leaky ReLU) is adopted,
\begin{align}
    f_{\text{leaky ReLU}}(t)=\bigg\{\begin{aligned}
        &t\quad\ \,t\geq0\\
        &at\quad t<0
    \end{aligned}\,,
\end{align}
with a small positive constant $a$.
This activation function is computationally efficient and widely used in practice.

Despite its conceptual simplicity, the perceptron model is restricted to learning linearly separable mappings \cite{Minsky.etal1969}.
To overcome this restriction, more complex models like multilayer perceptrons (MLPs) \cite{cybenko_approximation_1989} are invented, as illustrated in Fig.~\ref{fig:nn}(b).
An MLP consists of multiple fully connected layers, in which the scalar weights of the perceptron are generalized to weight matrices.
The layers in the middle do not serve as output, thus termed as hidden layers.
The forward propagation through the network is described by 
\begin{align}
    \vec{x}^{(i+1)} = f\bigl({\vec{W}^{(i)}} \trans \vec{x}^i\bigr)\,.
\end{align}
By stacking multiple nonlinear transformations, MLPs acquire the capacity to approximate highly complex functions.
The number and scale of the hidden layers determine the ability of approximation.
In principle, sufficiently large networks can approximate arbitrary continuous mappings.
However, increasing network depth and width significantly complicates the training and convergence.
This challenge is particularly pronounced in quantum state tomography, where the dimension of the Hilbert space grows exponentially with the number of qubits.
The networks are trained using the mean squared error loss and the Adam optimizer.
Further training details are available in the code repository \cite{code_NNQST}.

\subsection{The noise model and measurements}
The MLP-based approach is intrinsically data-driven and does not require explicit analytical models for either the noise process or the measurement apparatus.
Instead, the network adapts to a given experimental setting through supervised training on appropriately generated datasets.
As a result, both the noise acting on the quantum state and the choice of measurements may, in principle, be arbitrary.
The specific noise and measurement models introduced below, therefore, serve as representative examples rather than fundamental restrictions of the method.

While the noise may take any form, the noise model must be parameterized in order to perform numerical simulations.
Several commonly encountered channels are considered, including the depolarizing noise, bit-flip noise, phase-flip noise, and amplitude-damping noise.
These noise channels are applied sequentially to the ideal quantum state.
For each state and each channel, the noise strength is sampled uniformly from a predefined interval, thereby generating a dataset with continuously varying noise levels.
Through training on such data, the neural network learns to recognize and compensate for the associated noise in the measurement statistics.
Although neural networks generally exhibit good generalization, the accuracy may decline significantly outside the range of the training set.
Therefore, training with diverse noise structures and broader intensity ranges helps improve the model's applicability and robustness.

The set of measurements depends on the prior knowledge of the degrees of freedom of the system.
An informationally complete set of measurements, such as the symmetric informationally complete positive operator-valued measure (SIC POVM), is a good choice if there are no restrictions on the state.
In this work, simple local Pauli projective measurements are adopted as a practical alternative, i.e.,
\begin{align}
    & M=M_1\otimes M_2\otimes\cdots\otimes M_n\,,\nonumber\\
    & M_i\in \{\openone, \sigma_x, \sigma_y, \sigma_z\}\,.
\end{align}
excluding $\openone^{\otimes n}$, where $\sigma_{\{x,y,z\}}$ are the three Pauli matrices.
For $n$ qubits, it yields $4^n-1$ independent outcomes, matching the number of freedoms in the corresponding systems.
The resulting measurement set corresponds to a coarse-grained cube measurement \cite{Burgh.etal2008Choice}, which reduces the input dimensionality, thus allowing smaller networks, improved scalability, and accelerated training.

\section{Applications}\label{Sec:App}
In this section, the neural-network framework introduced above is assessed from structured pure states to general mixed states, allowing different aspects of the reconstruction performance to be examined.
All simulations are performed using the noise and measurement models described in the previous section.
These results illustrate both the characteristic behavior of the neural-network-based approach and its feasibility for multi-qubit systems.
The Python codes for quantum state tomography with neural network algorithms, with accompanying documentation and implementations, are available online \cite{code_NNQST}.

\subsection{Pure states}
The first task involves tomography of two classes of highly structured pure states: GHZ-like states and Dicke states.
The GHZ-like state is defined as:
\begin{align}
    \ket{\psi_{\text{GHZ-like}}} = \cos{\theta}\ket{00\cdots0}+\sin{\theta}\ket{11\cdots1}\,.
\end{align}
And the Dicke state writes:
\begin{align}
    \ket{\psi_{\text{Dicke}}} = \frac{1}{\sqrt{C_n^k}}\sum_l \cP_l\{\ket{1}^{\otimes k}\otimes\ket{0}^{\otimes(n-k)}\} \,,
\end{align}
where $\sum_l \cP_l\{\cdot\}$ denotes the sum over all possible permutations. 
Parameters $\theta$ and $k$ are uniformly sampled within their respective ranges.
The dataset comprises an equal mixture of these two classes of states, preventing the network from converging to a fixed output, also enabling an evaluation of the tomography performance across different types of states.

\begin{figure}
    \centering
    \includegraphics[width=0.95\linewidth]{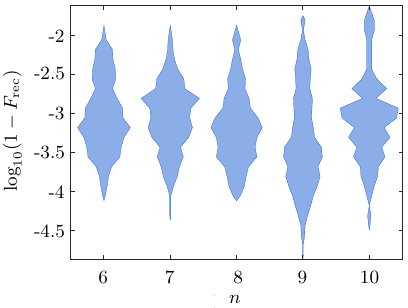}
    \caption{Tomography of six- to ten-qubit mixtures of GHZ-like and Dicke states.
    Among one thousand samples in each case, the fidelity of the noisy state satisfies  $0.75\leq F_{\text{noise}} \leq1$.
    Histogram width indicates sample counts at each infidelity $1-F_{\text{rec}}$.
    Most reconstructed states achieve infidelities below $10^{-3}$, with the worst-case infidelity below $0.025$.
    }
    \label{fig:6-10qubit}
\end{figure}

Pure states contain fewer degrees of freedom and therefore do not require informationally complete measurements for reconstruction.
Consequently, a randomly chosen subset of informationally complete measurements is enough.
Noise mitigation is highly effective in this typical scenario, as illustrated in Fig.~\ref{fig:6-10qubit}, taking one thousand random six- to ten-qubit systems each as an example.
In this test, the neural networks have three hidden layers of size of $100\cdot2^{\lceil\frac{n-4}{2}\rceil}$ each.
Note that for notational clarity, the subscripts ``ori'', ``noisy'', and ``rec'' denote the ideal noise-free state, the noise-affected state, and the state reconstructed by the neural network, respectively.
The average fidelity across a wide range of parameters and qubit numbers satisfies ${\bar{F} > 0.995}$.
Notably, no systematic degradation of performance is observed as the number of qubits increases, indicating favorable scalability of the neural-network-based approach.
Although there is a moderate spread in reconstruction fidelity, all reconstructed states in the numerical tests achieve fidelities of at least $0.975$.

Moreover, the number of measurement settings required for reconstructing pure states scales as $2^{n+1}$ for $n$ qubits.
This scaling is significantly smaller than that of an informationally complete set but matches the number of degrees of freedom in pure states.
These measurements are selected at random from the informationally complete set.
Although insufficient measurements may occasionally succeed for specific states due to randomness, reliable tomography across the entire state family generally requires sufficient information even for states with reduced degrees of freedom.

\subsection{Mixed states}
\begin{figure}[t]
    \centering
    \includegraphics[width=0.95\columnwidth]{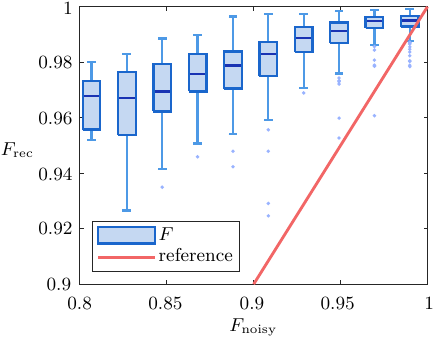}
    \caption{
    Fidelity comparison for random two-qubit mixed states.
    The diagonal line $F_{\mathrm{rec}} = F_{\mathrm{noisy}}$ indicates the absence of noise mitigation.
    Data points above this line demonstrate successful noise suppression by the neural network.
    The results confirm noise mitigation for almost all the cases tested. 
    For most states, the reconstructed fidelity exceeds the noisy fidelity, with only a small fraction exhibiting slight degradation due to statistical learning effects.
    }
    \label{fig:2qubit}
\end{figure}

To address more general scenarios, the state space is extended to arbitrary mixed states. 
In this subsection, two-qubit systems are considered as a representative test case, allowing for a detailed statistical characterization of the reconstruction performance.
In addition to fidelity, physically relevant quantities such as purity and entanglement negativity are examined.

The fidelity distribution under ideal conditions is depicted in Fig.~\ref {fig:2qubit}, where probabilities replace frequencies.
Noting that a neural network with three hidden layers of 100 neurons each is applied.
The diagonal line ${F_{\mathrm{rec}} = F_{\mathrm{noisy}}}$ serves as a reference corresponding to the absence of noise mitigation.
Data points located above this line indicate successful mitigation of noise in state tomography.
The results demonstrate that for the majority of states among one thousand samples, the reconstructed fidelity exceeds that of the noisy input, confirming the effectiveness of the proposed approach.

In contrast to the pure-state case, mixed states lack strong structural constraints imposed by the training data.
Consequently, the neural network must infer noise patterns solely from the statistical regularities present in the training data.
Then, the average fidelity achieved for mixed states is lower than that observed for pure states.

\begin{figure}[t]
    \centering
    \includegraphics[width=0.95\columnwidth]{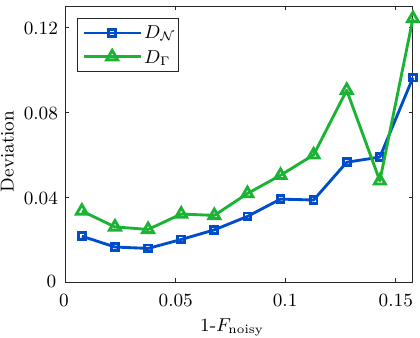}
    \caption{
    \label{fig:puriandneg}
    Deviations of negativity and purity for the reconstructed two-qubit random mixed states.
    They remain moderate over a wide range of reconstruction infidelities, which confirms that the noise mitigation is also effective in terms of purity and negativity.
    }
\end{figure}

Beyond fidelity, comprehensive characterization requires the analysis of purity ($\Gamma$) and negativity ($\cN$),
\begin{align}
    \Gamma &=\tr(\rho^2)\,,\nonumber\\
    \cN &=\frac{1}{2}\bigl(||\rho^\Gamma||_1 - 1\bigr)\,,
\end{align}
where $||\rho^\Gamma||_1$ is the trace norm of the partial transpose of $\rho$.
Negativity measures the degree of entanglement, while purity measures how much a state is mixed.
Figure~\ref{fig:puriandneg} displays the deviation of purity and negativity,
\begin{align}
    D_\Gamma &= |\Gamma_{\text{rec}} - \Gamma_{\text{ori}}|\,,\nonumber\\
    D_\cN &= |\cN_{\text{rec}} - \cN_{\text{ori}}| \,.
\end{align}
These deviations remain moderate over a range of infidelity values.
For instance, when infidelity is below $0.1$, the deviations are below $0.06$.
This further confirms the effectiveness of noise mitigation.
And, it indicates that the neural network not only improves fidelity but also preserves key physical properties of the underlying quantum states.

%
\begin{figure*}[t]
    \centering
    \includegraphics[width=1.95\columnwidth]{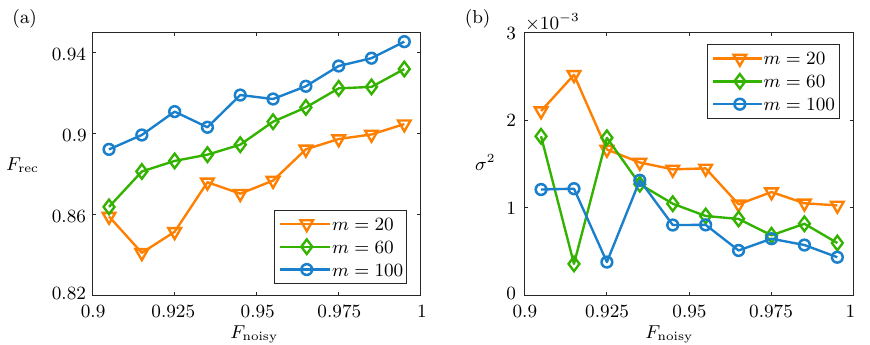}
    \caption{\label{fig:20-100}
    Reconstruction performance under simulated measurement statistics.
    The parameter $m$ is the average number of measurement repetitions.
    (a) Average fidelity as a function of $m$ and fidelity of noisy states. 
    (b) Variance of fidelity as a function of $m$ and fidelity of noisy states.
    Fidelity increases with reduced noise level and larger $m$, while
    variance generally decreases with lower noise and larger $m$.
    }
\end{figure*}

Furthermore, a closer inspection reveals a non-trivial dependence of fidelity and of deviations on the noise strength.
In particular, the highest average fidelity and lowest average deviations are achieved for states with a moderate level of noise, rather than for nearly noise-free states.
This appears counterintuitive, but follows directly from the network's fitting behavior.
Because the neural network infers noise from the training data, it tends to apply conservative denoising to optimize overall deviations.
As a result, nearly noise-free states may be partially interpreted as noisy, leading to slightly reduced fidelity in this regime.
The same conservative denoising behavior also governs mismatched scenarios.
When test noise differs from training noise, the network still applies the learned noise pattern, yielding lower fidelity than under matched conditions.

The reconstruction performance is further examined under simulated measurement statistics, where the parameter $m$ denotes the average number of repetitions for each measurement setting.
As shown in Fig.~\ref{fig:20-100}, increasing $m$ systematically enhances the reconstruction fidelity while simultaneously reducing its variance.
A similar improvement is observed when the physical noise level decreases.
This trend aligns with the expectation that reduced physical noise introduces smaller errors while more repetitions improve the precision of the data.

\section{Summary}\label{Sec:Sum}
This work demonstrates that neural-network-based quantum state tomography provides an effective and scalable strategy for mitigating unknown noise in quantum measurements.
By learning directly from data, our method enables high-fidelity reconstruction without requiring additional prior information about either the quantum state or the noise model.
Notably, for structured quantum states such as pure states, accurate reconstruction can be achieved with substantially fewer measurement settings.
As a result, this approach remains effective under limited measurement resources and naturally scales up to larger systems.
These characteristics significantly enhance its practicality for realistic experimental settings with constrained capabilities under noise.

On the other hand, the present framework assumes temporally stationary noise.
Therefore, extending the method to time-dependent or correlated noise represents an important and interesting direction for future research.
While multilayer perceptrons serve as a strong baseline, architectures capable of modeling temporal structure, such as recurrent neural networks, may offer improved performance in such scenarios.
Last but not least, incorporating physical constraints through physics-informed neural networks also constitutes a promising avenue for enhancing reconstruction accuracy and broadening applicability.

\acknowledgments
We are grateful to Ye-Chao Liu for helpful discussions.
This work was supported by the National Natural Science Foundation of China (Grants No.~12175014 and No.~92265115) and the National Key R\&D Program of China (Grant No.~2022YFA1404900).


%

\end{document}